\documentstyle[aps]{revtex}
\include{epsf}
\newcommand{\be}{\begin{equation}}
\newcommand{\ee}{\end{equation}}
\newcommand{\bea}{\begin{eqnarray}}
\newcommand{\eea}{\end{eqnarray}}
\newcommand{\beao}{\begin{eqnarray*}}
\newcommand{\eeao}{\end{eqnarray*}}
\newcommand{\e}{{\rm e}} 

\renewcommand{\d}{{\rm d}}

\newcommand{\Ref}[1]{(\ref{#1})}

\newcommand{\bc}{boundary conditions }

\newcommand{\gse}{ground state energy }
\newcommand{\hke}{heat kernel expansion }
\newcommand{\hkk}{heat kernel coefficient }
\newcommand{\hkks}{heat kernel coefficients }
\newcommand{\del}{\delta}
\newcommand{\uv}{ultraviolet }
\renewcommand{\em}{electromagnetic }

\renewcommand{\P}{{\cal P} }

\newcommand{\la}{\lambda}
\newcommand{\E}{{\cal E}}
\newcommand{\M}{{\cal M}}
\newcommand{\intd}{{\int \d\vec{x}}}
\newcommand{\sumn}{{\sum_{(n)}}}
\newcommand{\Mvariable}[1]{#1}

\def\beq{\begin{eqnarray}}
\def\eeq{\end{eqnarray}}
\newcommand{\nn}{\nonumber}

\newcommand{\sukeu}{\sum_{k=1}^\infty}

\newcommand{\sz}{z^{1/2}}
\newcommand{\zz}{(1+z^2)^{1/4}}

\newcommand{\pa}{\partial}

\newcommand{\cao}{{\cal O}}

\newcommand{\al}{\alpha}
\newcommand{\ep}{\epsilon}

\begin{document}

\draft

\title{On the ground state energy for a penetrable sphere and for a 
dielectric ball}
\author{{\sc  M. Bordag\thanks{e-mail:  Michael.Bordag@itp.uni-leipzig.de}, 
K. Kirsten\thanks{e-mail:  Klaus.Kirsten@itp.uni-leipzig.de} 
and D. Vassilevich}
\thanks{Alexander von Humboldt fellow. On leave from the Dept. of
Theoretical Physics, St.Petersburg University, 198904
St.Petersburg, Russia.
e-mail:  Dmitri.Vassilevich@itp.uni-leipzig.de}}
\address{ 
Universit\"at Leipzig, Fakult\"at f\"ur Physik und Geowissenschaften \\
Institut f\"ur Theoretische Physik\\
Augustusplatz 10/11, 04109 Leipzig, Germany
} 
\maketitle
\begin{abstract}

We analyse the ultraviolet divergencies in the ground state energy for
a penetrable sphere and a dielectric ball. We argue that for massless
fields subtraction of the ``empty space'' or the ``unbounded medium''
contribution is not enough to make the ground state energy finite
whenever the heat kernel coefficient $a_2$ is not zero. It turns out
that $a_2\ne 0$ for a penetrable sphere, a general dielectric
background and the dielectric ball. To our surprise, for more singular
configurations, as in the presence of sharp boundaries, the heat
kernel coefficients behave to some extend better than in the
corresponding smooth cases, making, for instance, the dilute
dielectric ball a well defined problem.
\end{abstract}


\section{Introduction}\label{Sec1}

The renewed interest in the problem of the 
\gse for the \em field in the presence of a dielectric body is 
triggered by the Schwinger suggestion \cite{Schwinger} that
the Casimir effect may serve as an
 explanation for the  sonoluminescence (for a review, see \cite{sono}). The
results by different groups are
controversally \cite{mn1,mn,visser}
and although there is a large number of papers on
this topic, basic issues are still to be clarified. As it will turn out, the
main difficulty is a proper formulation of the problem in the sense of finding
the right physical setup.

There are two ways to calculate the Casimir energy for a given configuration.
The first way consists of summing up the retarded Van der Waals forces between
individual molecules. The second way, which is studied in the present paper,
is to employ the technique of quantum field theory under the influence of
external conditions. 
(For a dilute dielectric ball the two methods have been shown to 
yield the same answer \cite{bmm} and in addition they are in agreement with 
a calculation of the Casimir energy treating the dielectric ball
as a perturbation \cite{gabriel}.)
This latter way is based on the evaluation of closed
vacuum loops with suitably modified propagators. Ultraviolet divergencies are
then removed by means of a renormalization procedure. While in the case of
conducting boundaries it is sufficient to subtract the empty space
contribution in order to make the vacuum energy finite, this is not the case
for penetrable boundary conditions, like 
the dielectric ball for instance. The main
aim of the present paper is to study the ultraviolet divergencies for the
electromagnetic field in the latter case.

{}From the early history of the Casimir energy calculations in a dielectric
using the quantum field theory framework we mention the papers
\cite{SDM,Mil,Candelas,BaKr}, and the more recent publications
\cite{BSS,BM}. More extensive bibliography can be found in \cite{Mil,BSS}
(see also the book by Bermann \cite{Berm}).

The problem can be formulated as follows. Consider a macroscopic body whose
\em properties are described by the permittivity $\ep (x)$, which may be
position dependent, i.e., which shows dispersion.  The function $\ep (x)$
enters the photon propagator as an external field 
respectively as a background.
Ultraviolet divergencies of the \gse of the \em field become functionals of
$\ep$.  In principle, it should be possible to interpret the \uv divergencies
inherent to the \gse in a way which yields a well defined, unique result. This
means that a classical model for the body must be found which allows to absorb
the divergent contributions (counter terms) into a redefinition of some
constants or some meaningful renormalisation condition must be found.

In a standard approach one uses the \hke in order to determine the counter
terms. In $(3+1)$ dimensions it is the coefficient $a_{2}$ which must be known
for a massless field, having in mind the \em field for instance. By now, it
is still unknown for a dielectric ball. The problem is that such a 
background is
not differentiable and the standard recursive formulas cannot be applied. On
the other hand,  the known results for manifolds with boundary (e.g. with
Dirichlet \bc ) do not apply because the dielectric ball results in some
penetrable \bc . Without attempting to solve this problem in general, we
calculate the relevant \hkks in two sufficiently characteristic examples,
first a spherical surface carrying a delta function potential ('delta sphere',
$V(r)\sim\delta(r-R)$) which can be reformulated in terms of matching
conditions at $r=R$, and second, the dielectric ball with constant permittivity
and permeability inside and outside. In both cases we discuss in parallel the
corresponding formulas for a smooth background.

Before actually calculating the \hkks we discuss the renormalization scheme in
general,
especially the problem of finding a normalization condition. We
argue that there is a natural condition for the massive field and show the
difficulties for the massless case caused by the conformal anomaly. 

Because of the contradicting results mentioned above we found it useful
to reconsider the quantization procedure for the Maxwell field in
a dielectric. In our previous paper \cite{bkv} the quantization in
terms of usual vector potentials $A_\mu$ was considered. We obsereved
no cancellation between ghosts and ``non-physical'' modes. Here we
re-examine this problem in terms of the so-called dual potentials.
We give a rigorous proof that for the spherically symmetric dielectric
ghost contribution can be neglected. We also show that the both 
approaches give identical results for the heat kernel coefficient
$a_2$ in the dilute approximation.

The paper is organized as follows. In the next section we discuss the
renormalization scheme, in sections 3 and 4 we calculate the \hkks for a
delta sphere and for the dielectric ball, the results are discussed in
section 5. Some formulas are given in the appendix.

\section{Renormalization and conformal anomaly}\label{Sec2}

Let us consider the \gse of a massive 
quantum field in the zeta-function regularisation,
\be\label{gse1}
\E_{0}=\frac{\mu^{2s}}{2} ~ \zeta_{\P}(s-\frac 1 2 ),
\ee
where
\be\label{zeta1}
\zeta_{\P}(s)=\sumn (\la_{(n)}+m^{2})^{-s}
\ee
is the zeta function of the operator $\P$
corresponding to a given dynamical system,
\be\label{op}
\P \phi_{(n)}(\vec{x})=\la_{(n)} \phi_{(n)}(\vec{x}).
\ee
The index $(n)$ includes discrete and continuous parts in dependence of the
problem considered. An arbitrary parameter $\mu$ has the dimension of a
mass. The zeta function is connected with the heat kernel by means of
\be\label{zhk}
\zeta (s)=\int_{0}^{\infty}\d t ~{t^{s-1}\over \Gamma (s)} K(t).
\ee
The first few terms (up to $n=2$ in (3+1) dimensions) of its asymptotic
expansion for $t\to 0$
\be\label{hke}
K(t)\sim {\e^{-t m^{2}}\over (4\pi t)^{3/2}}\sum\limits_{n}a_{n}t^{n}
\ee
yield the \uv divergencies of the \gse \Ref{gse1}. From that we define the
'divergent part' of the \gse as \cite{bekl}
 \bea\label{ediv} \E_{0}^{\rm div} &=& -\frac{m^4}{64\pi^2}\left(\frac 1 {s}
   +\ln\frac
   {4\mu^2}{m^2} -\frac 1 2 \right) a_0 -{m^{3}\over 24\pi^{3/2}}a_{1/2}\nn\\
 & &+\frac{m^2}{32\pi^2}\left(\frac 1 {s}
   +\ln\frac{4\mu^2}{m^2} -1\right) a_1+{m\over 16 \pi^{3/2}}a_{3/2}\\
 & &-\frac 1 {32\pi^2} \left( \frac 1 {s} +\ln \frac{4\mu^2}{m^2}-2\right)
 a_2. \nn \eea
Now the renormalised \gse can be defined as
\be\label{eren}
\E_{0}^{\rm ren}=\E_{0}-\E_{0}^{\rm div}
\ee
in the limit $s\to 0$. In general, the renormalisation is not unique as it is
obvious already from the presence of the parameter $\mu$. Therefor some
normalisation condition is necessary. 

In general, it depends on the considered physical problem how to handle this
non uniqueness. The simplest example is the Casimir effect between two
separate reflecting bodies with surfaces $\pa\M_{1}$ and $\pa\M_{2}$. In flat
space and without background potential the coefficients $a_{n}$ are integrals
over the surface and for $\pa\M_{1} \cap\pa\M_{2}=\emptyset$ they are
independent of the distance between the bodies. Therefor all divergent and
non unique contributions do not depend on the distance and a force as a
physical quantity can be defined. Note that this is sufficient for all current
experiments on Casimir force measurements. This applies 
especially to the case where $\pa\M_1$ and $\pa\M_2$ are 
nonintersecting spheres.
But it does not apply to the case
of a single sphere when asking for the pressure on its surface , for instance.

In the case of a massive field there is a natural normalisation 
condition. It
is the requirement that the renormalised \gse must vanish for the mass $m$
becoming large,
\be\label{normcond} \E_{0}^{\rm ren}\to 0~~\mbox{for}~~m\to\infty\,.  
\ee
Because the \hke is equivalent to the asymptotic expansion of the \gse for
large $m$ (provided there is no other dependence on $m$ than that in Eq.
\Ref{zeta1}), 
the definition
\Ref{ediv} of the divergent part through the \hkks is just equivalent to
\Ref{normcond}. This condition had been used in \cite{bekl} for the
calculation of the \gse of a massive scalar field with \bc on a
sphere.

Regrettably, this normalisation condition cannot be extended to the case of a
massless field. In general, for $m=0$, all dangerous contributions are
proportional to $a_{2}$. In case $a_{2}=0$ there is no divergence and no
arbitrariness at all and, hence, 
no need for a normalisation condition. But for
$a_{2}\ne 0$ the limit $m\to 0$ cannot be performed in $\E_{0}^{\rm ren}$
\Ref{eren} as can be seen quite easily.  For this we observe that the \gse
\Ref{gse1} can be expanded into a power series for small $m$ provided there
are no zero modes (we assume all $\la_{(n)}>0$)
\be\label{smallm}
\E_{0}
=\frac 1 2 \left(\sumn \la_{(n)}^{1/2-s} + m^{2}\sumn \la_{(n)}^{-1/2-s} + \dots\right)\,.
\ee
Now, by means of \Ref{eren} and \Ref{ediv} we subtract a contribution
containing $(a_{2}/16\pi^{2})\log m$. Therefor, the behaviour of the
renormalised \gse $\E_{0}^{\rm ren}$, eq. \Ref{eren}, for small $m$ is
\be\label{m2zero}
\E_{0}^{\rm ren} \raisebox{-8pt}{~$\stackrel{\sim}{m\to 0}$~} 
-{a_{2}\over 16\pi^{2}}\log m\,.
\ee
Consequently, there is no transition from the properly normalised \gse of a
massive field to the corresponding massless case as long as $a_{2}\ne 0$.

A more general procedure for handling the divergencies is the following. One
considers the quantum field in the background of some classical system. One
may think of quantised matter fields in the background of the classical
gravitational field or of quantum fluctuations around a classical solution.
The minimal structure of the background is determined by the \hkk $a_{n}$ with
$n=0,\dots,2$ which are certain functions respectively functionals of the
background. In fact, the structure of the energy of the classical system is
determined by $\E_{0}^{\rm div}$, eq. \Ref{ediv}. In writing
\be\label{eclass}
\E^{\rm class}=p a_{0}+\sigma a_{1/2}+F a_{1}+k a_{3/2}+h a_{2},
\ee
where $p,\sigma, F,k,h$ are the parameters of the classical energy, the
subtraction in \Ref{eren} can be interpreted as a redefinition of these
parameters. They remain undetermined and, hence, cannot be predicted from the
calculation of the \gse. In fact, by means of this, the \hkks give the physical
prediction which contributions are of quantum origin  and which are of
classical nature. 

Again, a remark must be added for the massless case. Consider, for example, a
scalar field with \bc on a sphere. The dependence of the eigenvalues on the
radius is $\la_{(n)}=\tilde{\la}_{(n)}/R^{2}$ where $\tilde{\la}_{(n)}$ are
numbers (depending on the kind of \bc). Hence the \gse can be written  as
\be\label{mless} 
\E_{0}={(R\mu)^{2s}\over 2R}~\tilde{\zeta}_{\P}(s-1/2),
\ee
where $\tilde{\zeta}_{\P}$ is the zeta function for $m=0$ and $R=1$. 
In case
$a_{2}\ne 0$, 
one has the expansion
\[\tilde{\zeta}(s-1/2)={-\tilde{a}_{2}\over 16\pi^{2}}{1\over s}+
\tilde{h}+O(s),
\]
where $\tilde{a}_{2}=a_{2}(R=1)$ and $\tilde{h}$ is some number which cannot
be calculated from the asymptotic expansion. The \gse becomes
\be\label{e1}
\E_{0}={-\tilde{a}_{2}\over 32\pi^{2}R}\left({1\over s}+\log
  (R\mu)^{2}\right)+{\tilde{h}\over R}.
\ee
The appearance of $\log R$ is just a result of the conformal anomaly. This was
clearly stated in \cite{bvw}, 
where the scaling behavior of the
Casimir energy was investigated
(see also \cite{stuart}). In order to setup the renormalisation scheme
one has to introduce the classical energy
\be\label{eclass2}
\E^{\rm class}={h\over R},
\ee
which allows to absorb the pole term into a redefinition of the parameter $h$.
However, the contribution $\tilde{h}/R$ in $\E_{0}$ which could be viewed as
the genuine result of the calculation of the \gse (sometime it is called the 
nonlocal contribution) cannot be distinguished from
the classical part and its calculation does not have a predictive power. Thus
the only outcome from the 
calculation of the \gse is the contribution containing $\log R$. This
consideration applies in the case $a_{2}\ne 0$. No renormalization procedure
and no classical system are required in case $a_{2}=0$. The \gse can be
calculated directly and becomes
\be\label{e2}
\E_{0}={\tilde{h}\over R}.
\ee

We remind that for the scalar field with Dirichlet or Robin \bc and for the
\em field with conductor \bc , the coefficient $a_{2}$ is the same except for
the sign for the interior and the exterior regions 
(the reason being that the extrinsic curvature has opposite signs
on both sides). Therefore,
when considering the \gse in the whole space, $a_{2}$ is zero and a result
like \Ref{e2} can be obtained. An example for this is just the well known
Casimir effect for a conducting sphere, where 'delicate' 
cancellations have
been observed in early papers already. 
In contrary, when considering only the interior
(or the exterior or an odd dimensional spacetime), one has $a_{2}\ne 0$ and
only the logarithmic contribution in \Ref{e1} is the part which it is
meaningful to calculate, i.e., it is sufficient to calculate $a_{2}$.

{}From this point of view, let us consider a 
conducting sphere of finite thickness.  Let a
massles scalar field be given in the regions 
($0\le r\le R_{1}) \ \cup \ 
(R_{2}\le r < \infty$), $R_1<R_2$, 
obeying Dirichlet \bc at $r=R_{1}$ and $r=R_{2}$. Let the
field be zero in between the two spherical shells having in mind an ideally
conducting medium there.  The \hkk is $\tilde{a}_{2}=-16\pi /315$. In zeta
functional regularization, the \gse for a single sphere 
is given by eq. \Ref{e1}. From that we obtain for two   
concentric spheres
\be \E_{0}={1 \over 630\pi} \ \left({1\over R_{1}}-{1\over R_{2}}\right) \ 
\left({1\over s}+\log (R\mu)^2\right)+{\tilde{h}_{1}\over
  R_{1}}+{\tilde{h}_{2}\over R_{2}}, \ee
where $\tilde{h}_{1}\ne\tilde{h}_{2}$ are the 'genuine', 
nonlocal contributions of the
\gse inside respectively outside a sphere of unit radius. 

Now, by writing
\[
{\tilde{h}_{1}\over R_{1}}+{\tilde{h}_{2}\over R_{2}}=
\tilde{h}_{1}\left({1\over R_{1}}-{1\over R_{2}}\right)+
{\tilde{h}_{1}+\tilde{h}_{2}\over R_{2}}
\]
we can define a classical energy
\be\label{eclass3}
\E^{\rm class}=h \left({1\over R_{1}}-{1\over R_{2}}\right)
\ee
where $h$ is some free parameter, and the renormalised energy by
(\ref{eren})
\[\E_{0}^{\rm ren}={\tilde{h}_{1}+\tilde{h}_{2}\over R_{2}},
\]
which delivers a meaningful expression for a pressure on the surface at
$r=R_{2}$. 

By writing
\[
{\tilde{h}_{1}\over R_{1}}+{\tilde{h}_{2}\over R_{2}}=
-\tilde{h}_{2}\left({1\over R_{1}}-{1\over R_{2}}\right)+
{\tilde{h}_{1}+\tilde{h}_{2}\over R_{1}}
\]
we get a similar result on the inner surface at $r=R_{1}$.  Thus, for
the given configuration, one is forced to introduce a classical model
with the energy (\ref{eclass3}) in order to obtain a renormalised
Casimir energy. However, although this procedure might look natural,
it isn't. So in the first case, the renormalized \gse is independent
on the inner radius which seems unlikely to be reasonable.

In summary, whereas for the massive field a unique 
normalized Casimir energy may be 
defined by considering the $m\to\infty$ behaviour, 
for the massless case, as seen for different situations, finite 
renormalizations remain undetermined as soon as the heat kernel coefficient
$a_2 \neq 0$.

\section{Penetrable spherical shell}\label{Sec3}

In the preceeding section we discussed the role of the coefficient $a_{2}$ for
the \gse, for instance the well known fact that there is a cancellation
between the two sides of a surface dividing the quantisation volume. Here we
show that this is a rather special case and that generally 
the renormalization ambiguity remains. We start from the known 
example of a
smooth background field $V(\vec{x})$. The operator $\P$ in \Ref{op} is
\be\label{pot}
\P=-\Delta+V(\vec{x})
\ee
and the corresponding \hkks are
\be\label{hkV}
a_{1}=-\intd \ V(\vec{x}),~~~~a_{2}=\frac 12 \intd \ V(\vec{x})^{2} .
\ee
The coefficient $a_{2}$ can vanish only in the trivial case of a vanishing
background potential. Consequently, the presence of $a_{2}$, i.e., of the
conformal anomaly, is a rather general case. The question arises how this is
connected with the cancellation of $a_{2}$ in the presence of boundaries. As
is known, there is no transition in the coefficients from a smooth
background to \bc. For instance, \bc lead to coefficients with half integer
numbers which are not present in \Ref{hkV}. In the following we consider a
singular background potential
\be\label{dpot}
V(r)={\al\over R} \ \delta(r-R) \ ,
\ee
which can be viewed as standing in the gap between a smooth background and
boundary conditions. 
The delta function potential \Ref{dpot} in the
operator $\P$, eq. \Ref{pot}, can be replaced by penetrable \bc, namely by the
matching conditions
\be\label{mc}
\phi_{|_{r=R-0}}=\phi_{|_{r=R+0}}, \qquad
\phi'_{|_{r=R-0}}-\phi'_{|_{r=R+0}}={2\al\over R}\phi_{|_{r=R}},
\ee
on the solutions of eq. \Ref{op} with $\P=-\Delta$. Clearly, the fomulas
\Ref{hkV} cannot be applied because $a_{2}$ would contain the delta
function squared. 

For $\al >0$ this potential is repulsive and in the formal limit
$\al\to\infty$ Dirichlet \bc at $r=R$ are restored. The Casimir force between
two planes carrying delta potentials had been calculated in \cite{bhr,hr} and
the similar problem for dissipative mirrors in \cite{jr}. 
In these cases
one had been interested in the distance dependent contributions only. 

For the calculation of the \gse and the \hkks we adopt the technique
developed in \cite{bk} for a smooth background potential. The starting point
is the expression of the \gse in zeta functional regularisation
\be\label{e0d} \E_{0}=-{\cos \pi s\over \pi}\sum\limits_{l=0}^{\infty}\nu
\int\limits_{m}^{\infty}\d k \ (k^{2}-m^{2})^{1/2-s}{\pa\over\pa k}\log
f_{l}(ik) \ee
with $\nu\equiv l+1/2$. Here, $f_{l}(ik)$ is the Jost function of the
scattering problem associated with the operator $\P$,
eq. \Ref{dpot}, on the imaginary
axis. It can be found by standard methods and reads
\be\label{jfd}
f_{l}(ik)=1+\al I_{\nu}(kR)K_{\nu}(kR),
\ee
where $I_{\nu}$ and $K_{\nu}$ are the modified Bessel functions. In  order to
calculate the \hkks it is sufficient to calculate the pole part of $\E_{0}$ in
$s=0$. For this reason we use the uniform asymptotic expansion of the Jost
function for $k\to\infty$, $\nu\to\infty$. From the corresponding asymptotic
expansions of the Bessel functions \cite{abra} we obtain
\be\label{lnfas} 
\log f_{l}^{\rm as}(ik)=\sum\limits_{n=1}^{3}\sum\limits_{i}X_{n,i}{t^{i}\over
  \nu^{n}}
\ee
with $t=1/\sqrt{1+(kR/\nu)^{2}}$ and $X_{1,1}=\al/2$, $X_{2,2}=\al^{2}/8$,
$X_{3,3}=\al/16+\al^{3}/24$, $X_{3,5}=-3\al/8$, $X_{3,7}=5\al/16$. Higher
terms of the expansion do not contribute to the pole.  Inserting $\log f^{\rm
  as}$ into the representation \Ref{e0d} we denote the corresponding part of
the 
\gse by $\E^{\rm as}$. There the integration over $k$ can be carried out
explicitely (e.g., using formula (31) in \cite{bk}). After that, the sum over
$l$ will be converted into two integrals by means of
\[\sum\limits_{l=0}^{\infty}f(l+\frac{1}{2})=\int\limits_{0}^{\infty}\d
\nu~f(\nu)-i\int\limits_{0}^{\infty}{\d \nu\over
  1+\e^{2\pi\nu}}\left(f(i\nu)-f(-i\nu)\right)
\]
and  $\E^{\rm as}$ splitted into two parts $\E^{\rm as}=\E^{\rm as,1}+\E^{\rm
  as,2}$ accordingly. In the first part the integral over $\nu$ can be
  carried out explicitely and we obtain
\beao
\E^{\rm as,1}&=&\frac{m^2}{32\pi^2}\left(\frac 1 {s}
   +\ln\frac{4\mu^2}{m^2} -1\right) (-4\pi\al R)+
{m\over 16 \pi^{3/2}}(\pi^{3/2}\al^{2})\nn\\
&&-\frac 1 {32\pi^2} \left( \frac 1 {s} +\ln \frac{4\mu^2}{m^2}-2\right)
 \left({\pi\al\over 3R}(1-2\al^{2})\right)\,.
\eeao
In $\E^{\rm as,2}$, from $X_{1,1}$ after some calculation a further pole
contribution results
\[\E^{\rm as,2}=-\frac 1 {32\pi^2} \left( \frac 1 {s} +\ln
  \frac{4\mu^2}{m^2}-2\right)\left(-\pi\al\over 3R\right)+O({1\over m}),
\]
where all other contributions vanish for $m\to\infty$ and hence do not
contribute to the considered \hkks .
Comparing these expressions with \Ref{ediv} we read off the coefficients 
\be\label{hkd}
a_{1/2}=0,~~ a_{1}= -4\pi \al R,~~ a_{3/2}=\pi^{3/2}\al^{2},~~ a_{2}=-{2\pi\over
  3}{\al^{3}\over R}
\ee
($a_{0}$ does not depend on the potential $V(r)$). 

{}From these formulas we see that the coefficients $a_{1/2}$ and $a_{1}$ are the
same as for a smooth background. In fact, in \Ref{hkV} $a_{1/2}=0$
and $a_{1}$ equals \Ref{hkd} when inserting $V(r)$,
eq. \Ref{dpot}. In contrast, $a_{3/2}$ in \Ref{hkd} has no counterpart in the
smooth case, which is not surprising as in the case of \bc coefficients with
half interger numbers are present. Finally, the coefficient $a_{2}$ in
\Ref{hkd} cannot vanish except for the trivial case $\al=0$ and it cannot be
obtained from $a_{2}$, eq. \Ref{hkV}, because of a squared delta function which
would appear. It is interesting to note that contributions linear in $\al$,
which are present in intermediate expressions $\E^{\rm as,1}$ and $\E^{\rm
  as,2}$, cancelled. We will see below that for the dielectric ball there is a
similar cancellation.
In order to renormalize the groundstate energy one needs to introduce the classical 
part ${\cal E}^{class}=ha_2$, see eq. (\ref{eclass}), of the energy and there is no condition
at hand to fix the free parameter $h$.
\section{Dielectric background}\label{Sec4}

There is a general interest in a careful consideration of the
quantisation procedure in a dielectric background. The \gse is the sum
over all states of the considered Hilbert space, physical as well as
unphysical ones. This is in opposite to the usual situation where
one has to consider expectation values in states with particles, i.e.,
in states of the physical subspace. Therefor the unphysical states,
including ghost states, must be considered carefully. A general
investigation whether the contributions of the ghosts and of the
unphysical photons cancel each other is still to be done. So a
reconsideration of the in general well known quantisation procedure of
QED in a dielectric background is in order. Thereby we consider two
quantisation procedures, first in terms of the electromagnetic
potentials and, secondly, in terms of the dual potentials. In fact,
the first case was already done in \cite{bkv}. We reconsider it here in
order to derive specific expressions in the spherically symmetric
case.

The action for the \em field in a medium with permittivity $\ep(\vec{x})$ and
permeability $\mu=1$ is
\be\label{em} 
S=\frac 12 \int d^4x (\epsilon (x) E^2 - B^2 ) .
\ee
First we consider the case of $\ep(\vec{x})$ being a sufficiently smooth
function so that the necessary number of derivatives exists. Furthermore, 
we assume that
$\ep(\vec{x})$ goes sufficiently fast to a constant value at spatial 
infinity
such that integrals appearing later on are well defined.
Volume divergencies still present have to be subtracted by hand. These 
terms are associated with the contributions of an unbounded medium, 
where $\ep$ everywhere takes its value at infinity.  
As is well known, for the
calculation of the \gse in gauge field theories care must be taken in order
not to
loose modes, ghost contributions for instance. This had been done in
\cite{bkv}. There, after introducing standard vector potentials, a gauge
fixing term, ghost fields, and taking into account the proper functional
measure, the path integral takes the   form

\bea\label{Z} 
Z&=&\int D\tilde A_i DA_0\ Dc\
D\bar c \exp \left\{i\int d^4x [ \frac 14 (2\epsilon (x) (\partial_0
\epsilon^{-1/2} \tilde A_i- \partial_i A_0)^2 \right.\nonumber \\ &\
& - (\partial_i \epsilon^{-1/2} \tilde A_k - \partial_k
\epsilon^{-1/2} \tilde A_i )^2) +{\cal L}_{gf}+ {\cal L}_{ghost}
]\bigg\}  
\eea
with $\tilde A_i =\sqrt \epsilon A_i$ and
\beao
{\cal L}_{gf}&=&-\frac 12 (\epsilon^{-1} \partial_i \epsilon^{1/2} \tilde A_i
- \epsilon \partial_0 A_0 )^2 , \nn \\
 {\cal L}_{ghost}&=& -\bar c (-\epsilon^{-1} \partial_i
\epsilon \partial_i + \epsilon \partial_0^2 )c \nonumber.
\eeao
The action for the electromagnetic field $A$ is then found to be
\beao & &\frac 12 \int d^4x[\epsilon (\partial_i A_0)^2
-\epsilon^2(\partial_0A_0)^2 +(\partial_0\tilde A_i)^2 \\
& &+\tilde A_i\epsilon^{-1/2} (\partial_j^2\delta_{ik}
-e_i\partial_k+\partial_ie_k-e_ ie_k)\epsilon^{-1/2} \tilde A_k], 
 \eeao
with the notation $e_i=\partial_i {\rm ln}\epsilon$, and later also 
$e_{ij}=\partial_i \partial_j {\rm ln}\epsilon$ and so on.

Now we are able to integrate over $A_0$, $\tilde A$ and the
ghosts. The resulting path integral reads after Wick rotation to the
Euclidean domain 
\[ 
Z=Z[A_0]Z[\tilde A]Z[\bar c,c] ,
\]
where the separate contributions are of the form: 
\beao 
Z[A_0]&=&{\det }^{-1/2} (-\partial_i \epsilon \partial_i -\epsilon^2
\partial_0^2 ), \nn \\ 
Z[\tilde A]&=& {\det }^{-1/2} \left (
-\frac 1\epsilon \partial_k^2\delta_{ij} -\partial_0^2\delta_{ij}
-{e_i}\epsilon\partial_j+{e_j}\epsilon\partial_i-
\frac 1\epsilon
(e_{ij}-e_ie_j) \right ), \\ 
Z[\bar c,c]&=& \det (-\epsilon^{-1}\partial_i \epsilon \partial_i - \epsilon
\partial_0^2 ). \nn
\eeao
For the functional determinants we use the integral representation
\[
 \log \det (L)= \int_0^\infty \frac {dt}t K(L;t) ,
\]
where
the heat kernel $K(L;t)$ for a second order elliptic operator $L$ is
\[ K(L;t)={\rm Tr} \exp (-tL)  .
\]

In  \cite{bkv}  we calculated the coefficients $a_n$
of the asymptotic expansion of the heat kernel 
\begin{equation}
{\rm Tr} (\exp (-tL))=(4\pi t)^{-2}\sum_{n=0}^\infty t^n
a_n (L)  .
\label{hke4}
\end{equation}
Note, that here $L$ is a four-dimensional operator. In eq.
(\ref{op}) above ${\cal P}$ is a three-dimensional operator.
This explains different powers of $4\pi t$ in (\ref{hke}) and
(\ref{hke4}).

Let us list the 
individual contributions to the different coefficients 
(summation  over
repeated indices is assumed). First 
\beao
a_0^{gh}&=& \int d^4 x \epsilon^{-1/2}  \nonumber \\
a_0^{A_{0}} &=&\int d^4x \epsilon^{-5/2} \nonumber \\
a_0^{\tilde A} &=&3\int d^4x \epsilon^{3/2}.
\eeao
For $a_1$ we obtain
\beao 
 a_1^{gh} &=&\int d^4x \epsilon^{-1/2} \ 
\left( -\frac 7{12} e_{ii} -\frac{13}{48} e_ie_{i}\right), \nonumber \\ 
 a_1^{A_{0}} &=&\int d^4x \epsilon^{-3/2} \ 
\left( -\frac 7{12} e_{ii} -\frac{13}{48} e_ie_{i}\right), \nonumber \\ 
 a_1^{\tilde A} &=& \int d^4x \epsilon^{1/2} \ 
\left( \frac 3{4} e_{ii} -\frac{1}{16} e_ie_{i}\right), \nonumber \\ 
\eeao
and for the coefficient $a_{2}$ the result is
\beao
 a_2^{gh} &=&\int d^4x \epsilon^{-1/2}
\frac 1{360}
( -33 e_{iijj} -18 e_ie_{ijj} -33 e_{ij}e_{ij} +\frac {237}{4}
e_{ii}e_{jj} \nonumber \\ \ &\ & +\frac {531}{8} e_{ij}e_ie_j +\frac
{33}{4} e_{ii}e_je_j +\frac {837}{64} e_ie_ie_je_j ),\nonumber \\
a_2^{A_0} &=&\int d^4x \epsilon^{-1/2} 
\frac 1{360}
( -27 e_{iijj} -60 e_ie_{ijj} -41 e_{ij}e_{ij}
+\frac {119}{4} e_{ii}e_{jj} \nonumber \\
 \ &\ & -\frac {91}{8}
e_{ij}e_ie_j +\frac {415}{8} e_{ii}e_je_j +\frac {4141}{64}
e_ie_ie_je_j ) +O(t^3) \}, \nonumber \\
a_2^{\tilde A} &=&  \int d^4x \epsilon^{-1/2} 
\frac 1{360}
( 33 e_{iijj} - 63 e_ie_{ijj} +231
e_{ij}e_{ij} -\frac {723}{4} e_{ii}e_{jj} \nonumber \\ \ &\ & -\frac
{1029}{4} e_{ij}e_ie_j +\frac {793}{8} e_{ii}e_je_j +\frac {4263}{16}
e_ie_ie_je_j )
\eeao
(in the first line of $a_2^{\tilde A} $ we corrected some tipos which 
appeared in \cite{bkv}).
 Because the
background, which we consider here, is static we can
drop the integration
over $x_{0}$. The contributions must be summed up according to
\be\label{polsum}
a_n=a_n^{A_0}+a_n^{\tilde A}-2a_n^{gh}.
\ee
The expression for $a_{2}$ appearing in this way represents the necessary
counterterm which is required to perform the renormalisation of the \gse of
the \em field. As it is seen, this is a quite complicated functional of
$\ep(\vec{x})$. It is interesting to consider the case of $\ep$ close to
unity (dilute dielectric medium). By means of $\ep=1+\delta$,
$e_{i}=\delta_{i}/(1+\delta)$, etc. we obtain
\beao
a_0 &=& \intd \ (2+3\delta +\frac{19} 4 \delta ^2 ) 
                    +O(\delta^{3}) \\
a_{1}&=&\intd \ (-\frac{11}{24} ) \ (\delta_{i})^{2} \ +O(\delta^{3})
\eeao
and
\be\label{a2ep}
a_{2}=\intd \ \frac{ 69}{720}\  (\delta_{ii})^{2} \ +O(\delta^{3}).
\ee
In $a_1$ and $a_2$ there are no terms of the 
first order, they cannot be present for symmetry and dimensional
reasons. 
(In some detail, the coefficient $a_1$ has the dimension of length, $a_2$ of 
$1/length$ and there is no possibility to have nonvanishing terms 
linear in $\delta$ which do not integrate away.)
But the terms of the 
second order are present. Moreover, they cannot
vanish (unless $\ep$ is globally constant).

{}From this one can conclude, that for a smooth $\ep(\vec{x})$ the \uv
divergencies in the \gse of the \em field are present generically and
can cancel at most for very specific choices of $\ep$ but not for
small $\ep -1$ as shown by \Ref{a2ep}.

The functional dependence of the needed counterterms is very involved 
and the classical model necessary for renormalization lacks physical 
intuition as well as motivation.

So we obtained the relevant \hkks for a $\ep( \vec x)$
which must be at least for times differentiable in order to deliver a
finite result when inserted into the equation for $a_2$. 
When restricting to a spherically symmetric situation, the 
quantization approach using dual potentials will provide several 
advantages. (For the general formalism of dual potentials see 
\cite{rohrlich}.) Probably most important is, that the calculation
of the relevant determinant can be reduced to the problem of two 
scalar fields. This procedure can be taken over to the 
non-smooth dielectric ball treated below and 
for that reason is described in some detail.

In this approach one starts with  
the action
\begin{equation}
S=-\frac 12 \int d^4x\left( \frac 1\epsilon 
\vec{D}^2 -\vec{B}^2 \right) ,
\label{action}
\end{equation} 
which generates the Maxwell equations in dielectric medium
with permittivity $\epsilon (x)$. Here
$\vec{D}=\epsilon \vec{E}$. 
Introducing the dual
potentials $C_\mu$: $\vec{D}=\vec{\nabla}\times \vec{C}$, 
$\vec{B}=\partial_t \vec{C} -
\vec{\nabla} C_0$,
one gets the 
canonical Poisson brackets,
\begin{equation}
\{ C_i (\vec{x},t), B^j(\vec{y},t)\} =\delta^j_i \delta
(\vec{x} -\vec{y}) ,\label{Poisson}
\end{equation}
thus $\vec B$ is the momentum conjugate to $\vec C$. 

One can rewrite the action (\ref{action}) in the canonical
first-order form,
\begin{equation}
S=\int d^4x \left( \vec{B}\partial_t\vec{C} +C_0
(\vec{\nabla}\vec{B}) -\frac 12 \vec{B}^2
-\frac 1{2\epsilon} \vec{D}^2 \right) .
\label{action1}
\end{equation}
$C_0$ plays the role of the Lagrange multiplier.

The theory (\ref{action1}) is then quantized according to the
general method \cite{FS}. The path integral measure reads
\begin{equation}
d\mu = DC_j DB^k DC_0 J_{FP} \delta (\chi ) \ ,
\label{dmu}
\end{equation}
where $\chi$ is  a gauge fixing condition, $J_{FP}$ is the
Faddeev--Popov determinant, $J_{FP}=\det \{ \chi ,
(\vec{\nabla} \vec{B} )\}$. For $\chi$ we choose the condition
\begin{equation}
\chi =\nabla_i C^i =0 \ .\label{chi}
\end{equation}
The Faddeev--Popov determinant becomes
\begin{equation}
J_{FP}=\det ( -\nabla^i \nabla_i ).
\label{FP}
\end{equation}
The ghost operator in (\ref{FP}) is non-elliptic, and, therefore,
the determinant (\ref{FP}) is ill-defined. Such operators must
be treated with great care in order not to break the gauge
invariance of the theory (see e.g. \cite{DV-gauge}). Fortunately,
in our case $J_{FP}$ does not depend on $\epsilon$ and can be
neglected alltogether. Moreover, at the expence of some technical
complications we can obtain an equivalent result in the Lorentz type
gauge $\nabla^\mu C_\mu =0$ where no such problems occur.

Now one can perform the intergrations over all fields arriving at
the following expression for the path integral,
\begin{equation}
Z[\epsilon ]=\int d\mu \exp (iS) =\int DC_\perp \exp (iS),
\label{Zep}
\end{equation}
where the subscript $\perp$ denotes
transversal 3-vectors, $\nabla_i C^i_\perp =0$.

Two substantial simplifications occur here compared to the quantization
in terms of non-dual vector potentials $A_\mu$ described before 
\cite{bkv}. First of all,
the gauge condition on $A_i$ \Ref{chi}, which removes the mixing between
$A_0$ and $A_i$, depends on $\epsilon$. As a consequence, the ghost
operator depends on $\epsilon$ as well and cannot be neglected. 
Also, in the case of non-dual potentials the term qudratic in momenta
is multiplied by $\epsilon^{-1}$. After momentum integration it leads
to a modification of the path integral measure.

Even though the path integral (\ref{Zep}) is Gaussian its direct
evaluation is not easy. The
corresponding determinant is to be computed on the space of transversal
vectors. To use standard heat kernel methods one must extend the
functional space to all unconstrained vector fields at the
expence of introducing a compensating scalar determinant. This
procedure is equivalent to introducing ghost fields. However,
if the function $\epsilon (x)$ is spherically symmetric, one
can reduce the determinant  arrising from 
(\ref{Zep}) to two scalar ones by 
making use of standard TE and TM modes. We turn to this case, now.

The transversality equation (\ref{chi}) can be solved by introducing
two independent scalar modes $\phi$ and $\psi$,
\begin{eqnarray}
&&C_r=r\tilde \Delta \psi , \nonumber \\
&&C_a=-\tilde\nabla_a \partial_r r \psi 
+r{\varepsilon_a}^b\partial_b \phi .\label{psiphi}
\end{eqnarray}
Here the coordinates $x^a$ denote angular variables, $\tilde \Delta$
and $\tilde \nabla$ are the Laplace operator and covariant derivative
on the two-sphere. $\varepsilon_{ab}$ is the two-dimensional Levi--Civita
tensor.

It is clear from (\ref{psiphi}) that the scalar harmonics with
zero angular momentum do not give rise to any transversal vector
fields. Such harmonics must be excluded from the path intergral. 

The normalization conditions for the scalar modes one can read
off from the integral
\begin{equation}
\int r^2dr\, d\Omega C_iC^i=
\int r^2dr\, d\Omega \left( \psi \Delta (\tilde \Delta r^2) \psi
+\phi (-\tilde \Delta r^2 )\phi \right),
\label{norm}
\end{equation}
where $d \Omega$ is the measure of the angular integration.

By substituting (\ref{psiphi}) in the action (\ref{action})
we obtain  
\begin{eqnarray}
&&S=\int r^2dr\, d\Omega \left[ \psi (r^2 \tilde \Delta )
\Delta \left( -\partial_0^2 +\frac 1\epsilon \Delta \right)
\psi \right. \nonumber \\
&&\qquad\ \left. +\phi (-r^2 \tilde \Delta )
\left( -\partial_0^2 + \left( \partial_r +\frac 1r \right)
\frac 1\epsilon \left( \partial_r +\frac 1r \right) 
+\tilde\Delta \right) \phi \right].
\label{act-new}
\end{eqnarray}

We can now evaluate the path integral (\ref{Zep}). Taking into
account the Jacobian factors appearing due to (\ref{norm}), one
arrives at the following expression
\begin{equation}
Z[\epsilon ]={{\det }'}^{-1/2} (L_\psi )
{{\det }'}^{-1/2} (L_\phi ), \label{Znew}
\end{equation}
with
\begin{eqnarray}
&&L_\psi =-\partial_0^2 +\frac 1\epsilon \Delta , \nonumber \\
&&L_\phi =-\partial_0^2 +\frac 1\epsilon \Delta +
\left( \partial_r \frac 1\epsilon \right)
\left( \partial_r +\frac 1r \right). \label{Ls}
\end{eqnarray}
The prime in (\ref{Znew}) reminds us to subtract zero angular
momentum fields. The operator $L_\psi$ is not hermitian.
However, its heat kernel coincides with that of the
hermitian operator $\epsilon^{-1/2} L_\psi \epsilon^{1/2}$.

The heat kernel coefficients for the operators $L_\psi$
and $L_\phi$ which {\it include} zero angular momentum
fields can be easily calculated by using the standard
methods \cite{Gilkey}. After continuation to the Euclidean
domain the oparators (\ref{Ls}) become elliptic operators
of the Laplace type. This means that they can be represented
in the form:
\begin{equation}
L=-(g^{\mu\nu}\partial_\mu\partial_\nu +a^\sigma \partial_\sigma +b)
\label{gen} \end{equation} 
where $g^{\mu\nu}$ is a symmetric tensor field which
plays the role of the metric. By introducing a connection
$\omega_\mu$, one can bring $L$ to the form:
\begin{equation} L=-(g^{\mu\nu}\nabla_\mu\nabla_ \nu +E) \label{gen1}
\end{equation} 
where $\nabla$ is a sum of the Riemannian covariant
derivative with respect to the metric $g$ and the connection $\omega$.
The explicit form of $\omega$ and $E$ is 
\begin{eqnarray}
\omega_\delta&=&\frac 12 g_{\nu\delta}(a^\nu
+g^{\mu\sigma}\Gamma_{\mu\sigma }^\nu ) \nonumber \\ E&=&b-g^{\mu\nu}
(\partial_\mu\omega_\nu + \omega_\mu \omega_\nu - \omega_\sigma
\Gamma^\sigma_{\mu\nu} ) \label{omega} \end{eqnarray} 
As usual, $\Gamma$ denotes the Christoffel connection. We must stress
here that the metric $g_{\mu\nu}$ is an auxilliary object. It does
not reflect any real geometry of the problem. The heat kernel
coeffcients can be expressed in terms of the fields introduced
above \cite{Gilkey}:
\begin{eqnarray} 
a_0&=& \int d^4x g^{1/2} \nonumber \\ 
a_1&=& \int d^4x g^{1/2}
(E+ \frac \tau{6}) \nonumber \\ 
a_2&=& \int d^4x g^{1/2} \frac 1{360} (60\tau E +180
E^2+30\Omega_{\mu\nu} \Omega^{\mu\nu} 
 +5\tau^2-2\rho^2+2R^2 ) \label{an}
\end{eqnarray} 
Here $R$, $\rho$ and $\tau$ are Riemann tensor, Ricci
tensor and scalar curvature of the metric $g$ respectively.  Semicolon
denotes covariant differentiation, $E_{;\mu}=\nabla_\mu E$.  All
indices are lowered and raised with the metric tensor. 
$\Omega$ is the field strength of the connection $\omega$, which is
zero for the operators (\ref{Ls}). This is the most effective,
though maybe a bit artificial way to calculate the leading terms
of the heat kernel expansion on a smooth background.

Metric and curvature tensors coincide for the operators
$L_\phi$ and $L_\psi$. In the dilute approximation they are:
\begin{eqnarray}
g_{ij}&=&\epsilon\delta_{ij} , \nonumber \\
{R^i}_{jkl}&=&\frac 12
(e_{jl}\delta_{ik}-e_{jk}\delta _{il} -e_{il}\delta_{kj}+e_{ik}\delta
_{lj})+\dots \nonumber \\ 
\rho_{jk}&=&-\frac 12 (e_{jk}+e_{pp}\delta_{jk}) +\dots \nonumber\\ 
\tau &=&-\frac 1{\epsilon}
2e_{pp} +\dots \, ,\nonumber
\end{eqnarray}
where we dropped the terms which do not contribute to the first
non-trivial order in the $(\epsilon -1)$ expansion. The functions
$E$ are different for $L_\psi$ and $L_\phi$:
\begin{equation}
E_\psi =\frac 14 e_{ii}+\dots ,\qquad
E_\phi =\frac 34 e_{rr} +\frac {e_r}{2r}+\dots \, .
\end{equation}

{}From these expressions we obtain 
 the conformal anomaly in the dilute approximation:
\begin{eqnarray}
&&{a_2}' [\phi ]=\int r^2dr\, d\Omega \frac 1{360}
\left[ \frac{39}2 \frac{\delta_r^2}{r^2} +\frac{129}4 \delta^2_{rr}
\right] , \label{aphi} \\
&&{a_2}' [\psi ]=\int r^2dr\, d\Omega \frac 1{360}
\left[ \frac{9}4 \delta^2_{ii}
\right] .\label{apsi}
\end{eqnarray}
Note, that the coefficient (\ref{aphi}) can not be represented
in the covariant form of the coefficient (\ref{apsi}). The prime
means that the zero angular momentum modes ($\phi^0$ and $\psi^0$)
are still to be subtracted. Introduce the fields $\bar \phi^0 =
r\phi^0$ and $\bar \psi^0 =r\psi^0$ which are square integrable
on the half-line with the unit weight. Acting on the fields
$\bar\psi^0$ and $\bar\phi^0$ the operators $L_\psi$ and
$L_\phi$ become
\begin{eqnarray}
&&L_\psi^0 =-\partial_0^2 +\frac 1\epsilon \partial_r^2 \nonumber \\
&&L_\phi^0 =-\partial_0^2 +\partial_r \frac 1\epsilon \partial_r
\label{Lzero}
\end{eqnarray}
Since the operators (\ref{Lzero}) are two-dimensional, the coefficient
before the zeroth power of proper time is given by $a_1$. Again, it can be
 calculated by the same methods of \cite{Gilkey}. The result is
\begin{equation}
a_1^0[\psi ]+a_1^0[\phi ]=
\int_0^\infty dr\, \left( -\frac 18 \delta_r^2 \right) =
\frac 1{4\pi}
\int r^2dr\, d\Omega \left( -\frac 18 \frac{\delta_r^2}{r^2} \right) .
\label{a10}
\end{equation}
The total coefficient $a_2$ after subtracting the zero angular
momentum modes reads
\begin{eqnarray}
a_2&=&{a_2}'[\psi ]+{a_2}'[\phi ]-
4\pi (a_1^0[\psi ]+a_1^0[\phi ] ) \nonumber \\
&=& \int r^2dr\, d\Omega \frac{69}
{720} \delta_{ii}^2 \ ,
\label{a2tot}
\end{eqnarray}
where the factor $4\pi$ appeared due to different normalization
conventions for the heat kernel coefficients in 2 and 4 dimensions.
We see, that the covariance of the result is restored. The
coefficient (\ref{a2tot}) coincides with the result of the calculations
in ordinary potentials, eq. (\ref{a2ep}), providing a check of 
our previous calculation.

Now, because the problem is reduced to two scalar ones we can turn to
the case of a dielectric ball of radius $R$ whose permittivity and
permeability are constant inside and outside
\be\label{epb}
\ep(r)=\ep_1 \ \Theta(R-r) +\ep_2 \ \Theta (r-R),
\ee
(and similar for $\mu (r)$). We remind that it is just the step
contained in this $\ep(r)$ why the above approach cannot be applied.

The most convenient form of the operators (\ref{Ls}) to
analyse the present case is
\begin{eqnarray}\label{L1}
&&L_\psi =-\partial_0^2 +\frac 1\epsilon \frac 1r \partial_r^2
r +\frac 1\epsilon \tilde\Delta \, ,\\\label{L2}
&&L_\phi =-\partial_0^2 + \frac 1r \partial_r \frac 1\epsilon
\partial_r r
 +\frac 1\epsilon \tilde\Delta \, .
\end{eqnarray}
The corresponding eigenfunctions, which can be divided into two
parts according to
\be\label{ansatzpsi}
\psi(r)=\psi_1(r)\Theta(R-r)+\psi_2(r)\Theta(r-R)
\ee
(similarly for $\phi $),
are determined by $\psi (r)$ 
and its first
derivative being continuous,
\be
r\psi \vert_+=r\psi \vert_-  ~~~~~~~~~~\qquad 
\partial_r r\psi \vert_+=\partial_r r\psi \vert_- 
\ee
and for $\phi$ we have 
\be
r\phi \vert_+=r\phi \vert_- ~~~~~\qquad 
\frac 1\epsilon \partial_r r\phi \vert_+=
\frac 1\epsilon \partial_r r\phi \vert_- \, ,
\ee
i.e., the function and a known combination of the first derivative
with $\ep$ must be continuous.  It is clear that these conditions are
just the same as known from the standard approach, see for example a
textbook on classical electrodynamics as \cite{stratton}.

Note that this situation is quite similar to the delta shell
\Ref{mc}. Just like in that case, to proceed further one has 
to define
the corresponding scattering problem and the corresponding Jost
functions. They turn out to be proportional to the expressions 
known for example
from \cite{mn1}
\bea\label{tem}
 \Delta _l ^{TE} (kR)&=& \sqrt{\ep_1 \mu_2}  s _l ' (k_1 R)  e _l
       (k_2 R) -\sqrt{\ep_2 \mu_1} s _l (k_1 R)
            e _l ' (k_2 R)  \nn \\
 \Delta _l ^{TM} (kR)&=& \sqrt{\ep_2 \mu_1} s _l ' (k_1 R)  e _l
       (k_2 R)-\sqrt{\ep_1\mu_2}   s _l (k_1 R)
             e _l ' (k_2 R)  
\eea
with the   notations
\be\label{sunde}
s _l (x) =\sqrt{\frac{\pi x}{2}} I_{\nu} (x) , \qquad
e_l (x) = \sqrt{\frac{2x}{\pi}} K_{\nu} (x) 
\ee
($\nu=l+1/2$).  Here, $\ep_{1,2}$ and $\mu_{1,2}$ are the permittivity and
permeability (we keep it in these formulas) inside respectively 
outside the ball.
Accordingly we use the notations $k_{1,2}=k\sqrt{\ep_{1,2}\mu_{1,2}}$.  These
expressions for $\Delta^{TE}$ and $\Delta^{TM}$ can be inserted into formula
\Ref{e0d} for the \gse whereby any multiplicative factors (even dependent on
$k$ as long as they are analytic near the real axis) by which $\Delta^{TE}$
and $\Delta^{TM}$ may differ from the Jost functions, drop out.  By means of
Eq. \Ref{zeta1} the corresponding zeta function then reads
\be \zeta (s) = \frac{\sin \pi s}{\pi} \sum_{l=1}^{\infty} (2l+1)
\int_0^\infty dk \ k^{-2s} \ \frac{\pa}{\pa k} \ln \left[ \Delta _l
^{TE} (kR) \Delta _l ^{TM} (kR) \right]. \label{eq2} 
\ee
 As will be seen clearly afterwards (eq. (\ref{heat0})), the Casimir
energy of the dielectric filling all space, and having permittivity
$\ep_2$ and permeability $\mu_2$, has already been subtracted.  This
is the translational invariant so called 'Minkowski space'
contribution.

It is useful to introduce the notation
\be
\Delta_{\rho , l}(ka) = \xi^\rho  s _l ' (k_1 a)  e _l
       (k_2 a) - s _l (k_1 a)
                e _l ' (k_2 a) \label{modcon}
\ee
with $\xi = \sqrt{\frac{\ep_1\mu_2}{\ep_2\mu_1} }$ and with the correspondence
$\Delta^{TE}_l\to \Delta _{1,l}$ and $\Delta^{TM} _l\to \Delta_{-1,l}$. In
passing from Eqs. (\ref{tem}) to (\ref{modcon}) only irrelevant factors have
been neglected.

As a result, after substituting $kR=z\nu$, the zeta function of TE
respectively TM modes read
\be
\zeta_\rho  (s) = \frac{\sin \pi s}{\pi} \sum_{l=1}^{\infty} (2l+1)
        \int_0^\infty dz \left(\frac{z\nu} a\right)^{-2s} 
              \frac{\pa}{\pa z}
              \ln \left[  \Delta _{\rho , l}(z\nu ) \right].\label{eq3}
\ee
In order to calculate the \hkks we need the uniform asymptotic expansion of
$\Delta_{\rho , l}$. It can be obtained just as in the preceeding section by
using the known uniform asymptotic expansions of the Bessel functions.  But
because the resulting expressions are more involved now, we introduce some
obvious abbreviations involving the Debye polynomials $u_k (t)$ 
and $v_k (t)$ \cite{abra},
\[\begin{array}{rclrcl}
A&=& \sukeu \frac{u_k (t)} {\nu^k} , \ \ \
B&=& \sukeu (-1)^k \frac{u_k (t)} {\nu^k} , \\[5pt]
C &=& \sukeu \frac{v_k (t)} {\nu^k} ,  \ \ \
D&=&\sukeu (-1)^k \frac{v_k (t)} {\nu^k} .  
\end{array}\]
In terms of these the asymptotic expansion of $s_l$, $e_l$, and their 
derivatives read
\bea
s_l (\nu z) &=& \frac 1 2 \frac{\sz}{\zz} e^{\nu \eta (z)} (1+A), \nn\\
e_l (\nu z) &=& \frac{\sz}{\zz} e^{-\nu \eta (z)} (1+B), \nn\\
s_l ' (\nu z) &=& \frac 1 {4\nu} \frac 1 {\sz \zz} 
                     e^{\nu \eta (z)} (1+A) \nn\\
            & &+\frac 1 2 \frac{\zz}{\sz} e^{\nu \eta (z)} (1+C),\nn\\
e_l' (\nu z) &=& \frac 1 {2\nu} \frac 1 {\sz \zz } 
                e^{-\nu \eta (z)} (1+B) \nn\\
          & &-\frac {\zz}{\sz}  e^{-\nu \eta (z)} (1+D). \nn
\eea
Before stating the asymptotic expansion of $\Delta_{\rho , l}$ some further
notation is necessary. First, the velocities of light are defined by $c_i =
1/\sqrt{\ep_i \mu_i}$ ($i=1,2$). Furthermore we use $t_i = 1/\sqrt{1+z_i^2}$,
$z_i = \sqrt{\ep_i \mu_i} z$, and we will write $A_i,B_i,C_i,D_i$ indicating
to which variable the contribution belongs. Then, after some calculation, we
get
\bea
\Delta_{\rho , l} &=& \frac 1 2 e^{\nu (\eta (z_1) - \eta (z_2))} 
         \frac {\xi^\rho c_1 t_2 +c_2 t_1}{\sqrt{c_1c_2t_1t_2}}\times\nn\\
  & &\left\{ 1+\frac{\sqrt{c_1c_2t_1t_2}}{\xi^\rho c_1 t_2 +c_2 t_1}
             \left[\xi^\rho \sqrt{\frac{c_1t_2}{c_2t_1}} \left[
               (C_1 +\frac{t_1}{2\nu} (1+A_1)) (1+B_2) +B_2 \right]
                            \right.\right.\nn\\
  & &\left.\left.+\sqrt{\frac{c_2t_1}{c_1t_2}}
                  \left[(D_2-\frac {t_2}{2\nu} (1+B_2))
                 (1+A_1) +A_1 \right] \right]\right\} .  \nn
\eea
The expansion for $\nu\to\infty$ of the logarithm entering  \Ref{eq3} reads
\[
\ln \Delta_{\rho,l}\sim 
\sum\limits_{n=-1,0,1,\dots}{D_{n,\rho}(z)\over \nu^{n}}
\]
with
\beao
D_{-1,\rho}(z)&=&\eta\left({z\over c_{1}}\right)-\eta\left({z\over c_{2}}\right),\\
D_{0,\rho}(z)&=& \ln \left( \frac 1 2 \frac{\xi^\rho c_1t_2+c_2t_1}
                { \sqrt{c_1c_2t_1t_2}} \right),
\eeao
the consecutive expressions are listed in the appendix. We define the
corresponding contributions to the zeta function $\zeta_{\rho}(s)$,
Eq. \Ref{eq3} to be
\be\label{}
A_{n,\rho}(s)= \frac{\sin \pi s}{\pi} \sum_{l=1}^{\infty} (2l+1)
        \int_0^\infty dz \left(\frac{z\nu} R\right)^{-2s}
              \frac{\pa}{\pa z} D_{n,\rho}(s)
\ee
($n=-1,0,1,2,3$). The contribution from $n=-1$ can be obtained doing the same
calculations as in \cite{bekl} for the massive scalar field in a ball with the
result
\bea
A_{-1,\rho } (s) &=& \frac{R^{2s}\Gamma \left(s-\frac 1 2\right)}
              {2\sqrt\pi \Gamma (s+1)} \zeta_H (2s-2;3/2) 
                     \left[\left(\frac 1 {c_1}\right)^{2s} -
                    \left(\frac 1 {c_2}\right)^{2s}\right]\label{eq7}.
\eea
The contribution following for the  residuum at $s=-1/2$  of the zeta
function is
\be
{\rm Res} A_{-1,\rho } (s=-1/2) = 
                \frac{127}{1920 \pi R} (c_1 - c_2) \label{eq8}.
\ee
The $z $-integral appearing in $A_{0,\rho}$ 
cannot be done by elementary means.
However the analytical structure around $s=-1/2$ is easily found. First, 
for $z\to 0$ the integrand behaves like $\cao ( z) $ 
at $s=-1/2$ and the integral
exists at the lower bound. Furthermore one can verify, that for $z\to \infty$
it behaves as $\cao ( z^{-2}) $ and the integral is well defined also at the 
upper integration limit. Thus $A_{0,\rho}$ is analytical around $s=-1/2$ 
and gives no contribution to the pole of the zeta function. 

Let us now come to the most delicate part of the calculation. As we will
see even the pole contribution will be, in general, a very involved 
integral which can only be analysed numerically. This will happen 
for the contribution arising from $D_{3,\rho} (z)$. But let us continue step 
by step. First, the angular momentum sum in the $A_{i,\rho}$ 
can be performed in terms of the 
Hurwitz zeta function, 
\be
A_{n,\rho } (s) = 2 \frac{\sin \pi s}{\pi} R^{2s}
            \zeta_H (2s+n-1;3/2) \int_0^\infty dz z^{-2s}
                       \frac{\pa}{\pa z} D_{n,\rho } (z) \label{eq9}
\ee
In the limit $z\to 0$ all $D_{n,\rho} (z)$ converge to a constant so that 
the integrals always exist at the lower bound. At the upper bound however
the integral might not be defined at $s=-1/2$. This happens for example 
for $D_{1,\rho} (z)$. Here one finds 
\be
D_{1,\rho} (z) = \frac 1 8 (c_1-c_2) 
    \frac 1 z +\cao \left( \frac 1 {z^3} \right)\nn
\ee
and the integral as it stands does not exist at $s=-1/2$. However, 
the analytical continuation is obtained by writing for $A_{1,\rho}$ 
instead of (\ref{eq9}) the following,
\bea
A_{1,\rho } (s) &=&  2 \frac{\sin \pi s}{\pi} R^{2s}
  \zeta_H (2s;3/2) \int_0^\infty dz z^{-2s}
   \left( \frac{\pa}{\pa z} D_{1,\rho} (z) 
               +\frac{c_1-c_2} 8 \frac 1 {(z+1)^2} \right) \nn\\
   & &-\frac 1 2 \frac{s( \sin \pi s )}{\sin (2\pi s)} R^{2s} (c_1-c_2) 
             \zeta_H (2s;3/2). \label{eq10}
\eea
Here, the asymptotics for $z\to \infty$ has been added and subtracted, 
the first term is now analytical around $s=-1/2$, the second one clearly 
contains the pole, namely
\bea
{\rm Res} A_{1,\rho} (s=-1/2) = -\frac{11}{192\pi R} (c_1-c_2).\label{eq11}
\eea
For $A_{2,\rho}$ similar arguments than the ones provided for 
$A_{0,\rho}$ show 
that there is no pole. Finally we are left with 
\bea
A_{3,\rho} (s) = 2 \frac{\sin \pi s}{\pi} R^{2s}
\zeta_H (2s+2;3/2) \int_0^\infty dz z^{-2s}
 \frac{\pa}{\pa z} D_{3,\rho } (z). \label{eq12}
\eea
Here at $s=-1/2$ the Hurwitz zeta function has already a pole such that 
the residue of $A_{3,\rho}$ is determined by the integral. In detail, 
after a partial integration one finds
\bea {\rm Res} A_{3,\rho} (s=-1/2) = \frac 1 {\pi R} \int_0^\infty \ dz \ 
D_{3,\rho }(z).
                  \label{eq13}
\eea
An integration of $D_{3,\rho} (z)$ in terms of elementary functions seems
impossible. For the case of $\mu_i =1 $ simplifications occur, but still a
complete analytical treatment seems only possible for $\rho = 1$, this is for
the TE modes.  For $\rho = -1$ the integral cannot be expressed in terms of
usual special functions. Instead we present the behaviour of the complete pole
(that is for the pole of the complete zeta function) as a function of the
velocities of light in Figure 1.

\begin{figure}\unitlength=1cm\begin{picture}(15,9)
\put(-1,-18){\epsfxsize=20cm \epsfbox{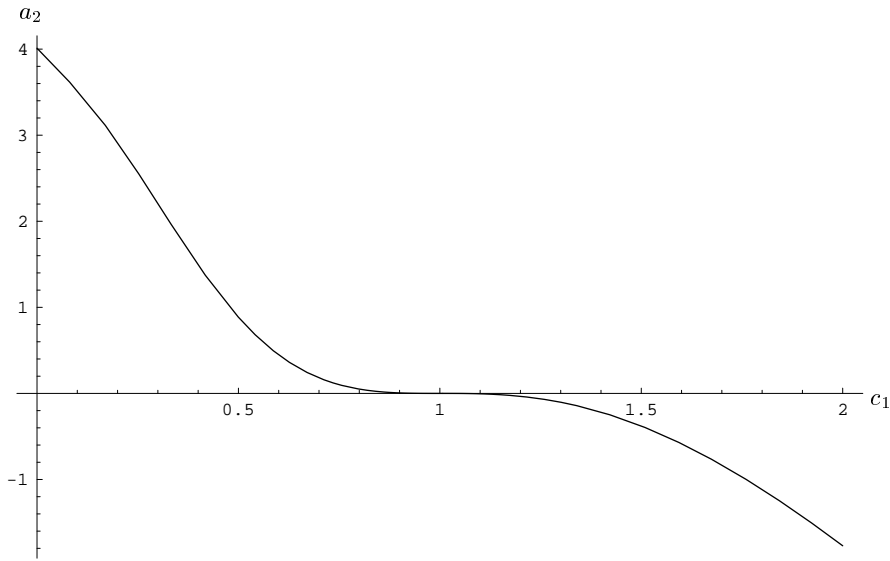} }
\end{picture}
\caption{The dependence of the \hkk $a_2$ on the speed of light $c_1=1/\sqrt{\ep_1}$ inside the ball while $c_2=1$ outside}
\end{figure}
In the limit of small differences of the velocities of light (dilute
dielectric ball) one might expand the residue in powers of $c_1-c_2$.
Surprisingly, the leading $\cao (c_1-c_2)$ pole coming from $A_{-1}$ and $A_1$
is exactly cancelled by the contribution coming from $A_3$, the second order is
zero so that we are left with a pole of the order $(c_1-c_2)^3$. 
This result is in agreement with \cite{bmm,BM}, where the calculation for a dilute 
ball was performed up to the order $(c_1 - c_2 ) ^2$ and a finite result was 
obtained in the zeta function scheme. Here we see that in the next order of the dilute 
approximation a pole appears.
In detail we
find
\bea
{\rm Res} \zeta (s=-1/2) = \frac {166}{5005\pi R} \frac{(c_1-c_2)^3}
                    {c_2^2}\label{eq14},
\eea
especially, for equal velocities of light in the interior and exterior 
region no pole is present. By means of \Ref{ediv}, the \hkk is
\be\label{a2}
a_{2}= -16 \pi^2 {\rm Res} \zeta (s=-1/2)=
-{2656 \pi\over 5005 R}{(c_{1}-c_{2})^{3}\over c_{2}^{2}}
+O((c_{1}-c_{2})^{4}),
\ee
a value quite close to $a_{2}$, eq. \Ref{hkd} for the delta sphere.

Let us mention that also for the $TE$ and $TM$ modes separately the divergent
term is of third order in the difference of the velocities of light. 

The peculiarity of this divergent term is, that it has the same $1/R$ dependence 
as the finite part of the result. This does not happen to quadratic order where 
independently of the scheme used \cite{bmm,gabriel,BM}, no such divergence appears.
Thus in contrast to the second order calculation, in the next order one must raise 
again the question how to fix the finite part of the Casimir energy. Until this 
question is not answered, the calculation of finite parts in higher orders of the 
dilute approximation makes no sense. The situation is even more involved for smooth 
dielectrics as seen in eq. (\ref{a2ep}) and as discussed further in the Conclusions.

However, before we come to the summary of our main results let us 
comment on other possible regularization schemes in order to make 
contact with the results of \cite{mn,bmm,gabriel}. 
It is known, that the zeta function 
scheme often gives finite answers whereas using 
other techniques divergences show up. 
An example is the dilute 
ball to second order in $(c_1-c_2)$ where the zeta-function 
method yields a finite result 
\cite{BM} whereas cutoff functions of different types 
yield poles proportional to the volume and surface for example. These divergent 
terms are also easily recovered by our calculation. There are well established 
connections between zeta function and cutoff regularization of Casimir energies.
Using a frequency cutoff $\lambda$ the divergent terms for $\lambda \to 0$ are 
\cite{gabimariel}
\beq
E_C^{div} = \frac{3a_0}{2\pi^2 \lambda^4} + \frac{a_{1/2}}{4\pi^{3/2}\lambda^3}
+\frac{a_1}{8\pi^2\lambda^2} +\frac{a_2}{16\pi^2} \ln \lambda \label{cutoff}
\eeq
with the heat-kernel coefficients $a_i$ of the problem at hand. These are immediately 
determined by the formulas of $A_{i,\rho}$ already presented. One gets 
\beq
a_0 &=& \frac 8 3 \pi R^3 \left[ \left( \frac 1 {c_1} \right)^3 -\left( \frac 1 {c_2}
\right)^3 \right], \label{heat0}\nn\\
a_{1/2} &=& -2R^2 \pi^{3/2} \frac{ (c_1^2 -c_2^2 ) ^2}{c_1^2 c_2^2 (c_1^2 +c_2^2)}
\nn\\
a_1 &=& -\frac{22} 3 \pi R \left( \frac 1 {c_1} -\frac 1{c_2} \right) 
  +8\pi R \int_0^{\infty} dy \,\, \frac 1 y 
\frac{\partial}{\partial y} \left[ D_{1,1} (y) +D_{1,-1} (y) \right] \nn\\
a_{3/2} &=& \pi^{3/2} \frac{ (c_1^2 -c_2^2 ) ^2 } { (c_1^2 +c_2^2 )^2 }\nn
\eeq
and $a_2$ is the sum of eqs. (\ref{eq8}), (\ref{eq11}) and (\ref{eq13}). It is clearly 
seen, that volume, surface and more divergencies are present. Using an expansion in 
$\del = c_1 -c_2$, one finds explicitly 
\beq
{\cal E}_0 &=& E_C^{div} +{\cal E}_0^{ren} \nn\\
&=& -\del \frac{9} {2\pi^2 c_2^4} \frac{V}{\lambda ^4} +
   \delta ^2 \left( \frac{18}{\pi^2 c_2^5} \frac V {\lambda^4} -\frac 1 {4\pi 
c_2^4} \frac S {\lambda^3} +\frac{23} {1536\pi} \frac 1 R\right) +{\cal O}  
(\delta ^3) ,\nn
\eeq
an expansion in parallel with Barton \cite{gabriel}, and where 
${\cal E}_0^{ren}$ has been taken from that reference. 
Let us mention that the factors of the divergent pieces
can not be compared to 
\cite{gabriel} because there a wave number cutoff has been used.
The fact that there is no $1/\lambda$ term is generic in this cutoff 
definition \cite{gabimariel}. In contrast, the reason for $1/\lambda^2$  
not being there is a result of cancellations between the $TE$ and $TM$ 
modes.
In the approximation given, volume and 
surface divergencies are absorbed by counterterms and 
a unique finite result is obtained. However, let us repeat and emphasize that 
at third order a term $(\delta ^3 / R ) \ln \lambda$ appears and renormalization and 
normalization of this term remains unclear.

\section{Conclusions}\label{Sec5}

We have calculated the lowest \hkks for the delta sphere and for the dielectric
ball. For a massless field, it is the coefficient $a_{2}$ which determines
the interpretation of the \gse. In both cases it does not vanish. For a delta
sphere it is the simple expression \Ref{hkd}, for the dielectric ball it
is an involved expression, including the integral \Ref{eq13}. 
Therefore, just subtracting the ``empty space'' or 
``unbounded medium'' contribution is
clearly insufficient to make the ground state energy finite.
At this point the conventional field theoretical approach fails to
give a unique result. It must be supplemented by an extra normalization
condition, which is not at hand by now.

A common feature of the two cases considered here is that for small
coupling to the background, i.e., for small $\al$ in \Ref{dpot} or
$\ep\to 1$ in \Ref{epb}, the lowest two orders vanish ( contrary to
the third and higher orders). Thereby nontrivial cancellations appear
in intermediate expressions.  Therefore, the good news is that
calculations in the leading orders of the dilute dielectric ball
approximation are physically relevant and give a unique result.  It
seems to be true for all penetrable \bc. This observation is supported
by an example of the paper \cite{mn}, where Casimir-Polder forces
between the particles constituting a dielectric sphere, had been
summed up over the volume.  In fact, the $r^{-7}$ potential had been
integrated using dimensional regularization. It turned out, that the
analytic continuation to the physical dimension yields a unique and
finite value.  In general, one expects the Casimir-Polder forces to be
equivalent to the \gse. Because these forces are nonadditive, the
result found in \cite{mn} should have relevance to a dilute dielectric
medium and in fact the identity of the van der Waals force and the
Casimir effect has been shown in \cite{bmm}.  Thus to the leading
non-vanishing order the situation is quite clear now for the dilute
dielectric ball, but, as we have described in detail at the end of
section 4 in higher orders the situation remains unclear. Even worse,
if we consider a smoothly varying dielectric background, we have seen
that already in the second order an ambiguity proportional to $a_2$,
eq. (\ref{a2ep}), remained. If one is going to calculate the Casimir
energy by summing up the van der Waals energies between the molecules
that make up the medium taking into account that $\epsilon$ and $\mu$
may vary in space, we expect that in the scheme used in \cite{mn} in
contrast to the dilute calculation divergencies or at last a
logarithmic contribution delivering a nonuniquness will show up when
the regularization is removed.

So for the case of nonconstant dielectricum the way how to fix a unique 
finite value of the Casimir energy still remains unclear
and further investigations will be necessary to shed more light into 
this very basic question.

\section*{Acknowledgments}
DV thanks the Alexander von Humboldt foundation and the
Russian Foundation for Fundamental Research, grant
97-01-01186, for financial support.
KK has been supported by the DFG under contract number Bo1112/4-2.

\appendix
\section{ Asymptotic functions $D_n (z)$}
In this appendix we give a list of the functions used in order to 
get the analytical properties of the relevant zeta function. We have,
\bea
\lefteqn{
D_{1,\rho} (z) = \frac 1 {24\,
    \left( \Mvariable{c_2}\,\Mvariable{t_1} +
       \Mvariable{c_1}\,\Mvariable{t_2}\,\xi^\rho 
          \right) }\times\label{eqa1}   }  \\
   & &  \left\{  - \Mvariable{c_2}\,\Mvariable{t_1}\,
        \left( -3\,\Mvariable{t_1} + 5\,{{\Mvariable{t_1}}^3} + 
          3\,\Mvariable{t_2} + 7\,{{\Mvariable{t_2}}^3} \right)    + 
     \Mvariable{c_1}\,\Mvariable{t_2}\,
      \left( 3\,\Mvariable{t_1} + 7\,{{\Mvariable{t_1}}^3} - 
        3\,\Mvariable{t_2} + 5\,{{\Mvariable{t_2}}^3} 
          \right) \,\xi^\rho \right\}, \nn
\eea
\bea
\lefteqn{
D_{2,\rho} (z) = \frac 1  {16\,\left( \Mvariable{c_2}\,\Mvariable{t_1} +
          \Mvariable{c_1}\,\Mvariable{t_2}\,\xi^\rho \right)^2} 
            \times \label{eqa2}      }   \\
    & &\left\{ \Mvariable{c_2}^2\,\Mvariable{t_1}^2\,
      \left( \Mvariable{t_1}^2 - 6\,\Mvariable{t_1}^4 + 
        5\,\Mvariable{t_1}^6 - \Mvariable{t_2}^2 + 
        6\,\Mvariable{t_2}^4 - 7\,\Mvariable{t_2}^6 \right)\right. \nn\\
   & & \left.+ 
     4\,\Mvariable{c_1}\,\Mvariable{c_2}\,\Mvariable{t_1}^4\,
     \Mvariable{t_2}^4\,\xi^\rho + 
     \Mvariable{c_1}^2\,\Mvariable{t_2}^2\,
      \left( -\Mvariable{t_1}^2 + 6\,\Mvariable{t_1}^4 - 
        7\,\Mvariable{t_1}^6 + \Mvariable{t_2}^2 - 
        6\,\Mvariable{t_2}^4 + 5\,\Mvariable{t_2}^6 \right)
        \,\xi^{2\rho} \right\} ,  \nn
\eea
\bea
\lefteqn{
D_{3,\rho} (z) =\frac 1 
     {5760\,
     \left( \Mvariable{c_2}\,\Mvariable{t_1} +
          \Mvariable{c_1}\,\Mvariable{t_2}\,\xi^\rho 
         \right) ^3} \label{eqa3}     }   \\
    & &
    \left\{   \Mvariable{c_2}^3\,\Mvariable{t_1}^3\,
      \left( 375\,\Mvariable{t_1}^3 - 4779\,\Mvariable{t_1}^5 + 
        9945\,\Mvariable{t_1}^7 - 5525\,\Mvariable{t_1}^9 \right.\right. \nn\\
   & &\left.+ 
        \Mvariable{t_2}^3\,\left( 345 - 4941\,\Mvariable{t_2}^2 + 
           11655\,\Mvariable{t_2}^4 - 7315\,\Mvariable{t_2}^6 \right) 
         \right)  \nn\\
    & &- 3\,\Mvariable{c_1}\,\Mvariable{c_2}^2\,
      \Mvariable{t_1}^2\,\Mvariable{t_2}\,
      \left( 1659\,\Mvariable{t_1}^5 - 3465\,\Mvariable{t_1}^7 + 
        1925\,\Mvariable{t_1}^9 + 
        120\,\Mvariable{t_1}^2\,\Mvariable{t_2}^3 - 
        720\,\Mvariable{t_1}^4\,\Mvariable{t_2}^3 + 
        600\,\Mvariable{t_1}^6\,\Mvariable{t_2}^3 \right.\nn\\
    & & \left. - 
        15\,\Mvariable{t_1}^3\,
         \left( 9 + 8\,\Mvariable{t_2}^2 - 48\,\Mvariable{t_2}^4 + 
           56\,\Mvariable{t_2}^6 \right)  + 
       \Mvariable{t_2}^3\,\left( -105 + 1581\,\Mvariable{t_2}^2 - 
           3735\,\Mvariable{t_2}^4 + 2275\,\Mvariable{t_2}^6 \right) 
         \right) \,\xi^\rho \nn\\
    & &   + 3\,\Mvariable{c_1}^2\,\Mvariable{c_2}\,
      \Mvariable{t_1}\,\Mvariable{t_2}^2\,
      \left( 1581\,\Mvariable{t_1}^5 - 3735\,\Mvariable{t_1}^7 + 
        2275\,\Mvariable{t_1}^9 - 
        120\,\Mvariable{t_1}^2\,\Mvariable{t_2}^3 + 
        720\,\Mvariable{t_1}^4\,\Mvariable{t_2}^3 - 
        840\,\Mvariable{t_1}^6\,\Mvariable{t_2}^3 \right.\nn\\
   & & \left.+ 
        15\,\Mvariable{t_1}^3\,
         \left( -7 + 8\,\Mvariable{t_2}^2 - 48\,\Mvariable{t_2}^4 + 
           40\,\Mvariable{t_2}^6 \right)  + 
        \Mvariable{t_2}^3\,\left( -135 + 1659\,\Mvariable{t_2}^2 - 
           3465\,\Mvariable{t_2}^4 + 1925\,\Mvariable{t_2}^6 \right) 
         \right) \,\xi^{2\rho} \nn\\
   & &     + \Mvariable{c_1}^3\,\Mvariable{t_2}^3\,
      \left( -345\,\Mvariable{t_1}^3 + 4941\,\Mvariable{t_1}^5 - 
        11655\,\Mvariable{t_1}^7 + 7315\,\Mvariable{t_1}^9 \right.\nn\\
   & &\left.\left.+ 
       \Mvariable{t_2}^3\,\left( -375 + 4779\,\Mvariable{t_2}^2 - 
           9945\,\Mvariable{t_2}^4 + 5525\,\Mvariable{t_2}^6 \right) 
         \right) \,\xi^{3\rho} \right\}  ,  \nn
\eea

\end{document}